\documentclass[12pt,times]{article}
\usepackage{times}

\def\ho:{\hbox{$H_0$:}}
\def\ha:{\hbox{$H_{\negthinspace A}$:}}
\def\ho{\hbox{$H_0$}}
\def\ha{\hbox{$H_{\negthinspace A}$}}

\newcommand{\comment}[1]{}

\def\boxit#1{\vbox{\hrule\hbox{\vrule\kern6pt
          \vbox{\kern6pt#1\kern6pt}\kern6pt\vrule}\hrule}}

\def\bse{\begin{eqnarray*}}
\def\ese{\end{eqnarray*}}

\def\beq{\begin{eqnarray}}
\def\eeq{\end{eqnarray}}

\def\bit{\begin{itemize}}
\def\eit{\end{itemize}}

\def\frac#1#2{{#1\over#2}}

\def\wh{\widehat}

\newcommand{\bma}[1]{\mbox{\boldmath $#1$}}

\newcommand{\bmu}{ {\bma{\mu}} }

\newcommand{\bSigma}{ {\bma{\Sigma}} }

\usepackage{graphicx}

\begin{document}

\newpage
\thispagestyle{empty}  
\baselineskip=30pt {\ } \vskip-1in

\begin{center}
 {\bf Estimating animal densities and home range in regions with irregular
boundaries and holes: a lattice-based alternative to the kernel
density estimator} \quad
\end{center}
 \begin{center}
{\LARGE {\bf }}
 \end{center}

\vskip 10mm \baselineskip=10pt
\begin{center}
Ronald P. Barry\par Corresponding Author\par Department of
Mathematics and Statistics
\par University of Alaska Fairbanks
\par rpbarry@alaska.edu\par
Fairbanks, AK 99775, USA\par Telephone:  (907) - 474 - 7226 \par
Fax: (907) 474 - 5394
\par
\end{center}

\vskip 6mm \baselineskip=10pt
\begin{center}
Julie McIntyre \par Department of Mathematics and Statistics \par
University of Alaska Fairbanks
\par jpmcintyre@alaska.edu\par
Fairbanks, AK 99775, USA \par
\end{center}

 \vskip 1.25in
 \baselineskip=14pt
 \noindent
{\bf Abstract}  Density estimates based on point processes are often
restrained to regions with irregular boundaries or holes.  We
propose a density estimator, the lattice-based density estimator,
which produces reasonable density estimates under these
circumstances.  The estimation process starts with overlaying the
region with nodes, linking these together in a lattice and then
computing the density of random walks of length k on the lattice. We
use an approximation to the unbiased crossvalidation criterion to
find the optimal walk length k.  The technique is illustrated using
walleye (Sander vitreus) radiotelemetry relocations in Lake Monroe,
Indiana.  We also use simulation to compare the technique to the
traditional kernel density estimate in the situation where there are
no significant boundary effects.

\vskip .25in \baselineskip=24pt \noindent {\bf Keywords:} Diffusion,
Point Process, Kernel Density Estimation, Intensity Functions,
utilization distribution, home range

\baselineskip=24pt
\newpage

\pagenumbering{arabic} 

\baselineskip=24pt

\section{Introduction}

Kernel density estimation is commonly used to estimate home ranges
and utilization distributions of fish or wildlife (for instance, see
Worton 1989).  A significant problem with kernel estimators,
however, is that they do not respect irregular boundaries or holes
in regions.  In estimating the home range of fish in a lake, for
example, a kernel density estimator will place positive density
along the shoreline or on islands within the lake.

A typical approach to remedying this problem is first to compute the
estimator as if there were no boundaries, then to clip off
inaccessible regions after the fact, and finally renormalize the
density.  But this solution is less than ideal.  For instance, if
there is a high density of fish in one lake and a second, closely
located lake has no observed fish, the home range might end up
including part of the apparently empty lake.  One approach to this
problem is the use of local convex hulls (Getz et al 2007; Getz and
Wilmers 2004; Ryan et al. 2006).

In this paper we suggest an estimator of density based on an
approximation to Brownian motion.  Random walks, originating from
each observation, are restrained to remain within the boundaries.
This would be analogous to adding a quantity of dye to each location
where a fish or animal was observed, then allowing the dye to
diffuse outward.  A density map based on the concentration of the
dye at different times would result in a density estimator that is
faithful to the boundaries of the region.

Our approach starts with a polygon that represents the region of
interest.  The polygon is filled with a grid of nodes, and a
neighbor relationship is defined on the nodes to create a lattice in
the sense of spatial lattice models (for instance, Cressie 1993, pp.
383 ff). The estimated density is derived from all length $k$ random
walks originating from the nodes where fish or other animals were
located. The length, $k$, of the random walk controls the smoothness
of the resulting estimate.  We propose a crossvalidation  approach
to select the optimal value of this smoothing parameter.

This density estimator we present has a number of desirable
properties that make it preferable to existing methods. For
instance, since the grid of nodes fills the polygonal region, the
resulting density estimator will automatically give zero density
outside the boundary.  Where observations occur in a restricted part
of the region, where nodes have fewer neighbors, the estimated
density will be higher, as expected.  It is also straightforward to
remove nodes within the region or links between nodes, accounting
for holes such as islands in lakes and boundaries such as causeways,
fences, etc.  The lattice-based estimator uses a neighbor
relationship betweens nodes filling a region, in contrast to the
network-based kernel estimator of Downs and Horner (2007) which
links observations into a network and uses kernel smoothing on the
resulting network.  The estimator is computationally fast, and we
have written a set of functions in R (R core development team 2009)
to implement it.

In this paper we  first describe the method, the {\it lattice-based
density estimator}, used to produce density maps.  We then present
 a simple example that illustrates the computations in
detail. Following this, we compare our method to a standard kernel
density estimation method for a two dimensional point processes.
Finally, we  apply the method to estimate the density of
walleye ({\it Sander vitreus}) in Lake Monroe, Indiana.

\section{Theory/Calculations}

\subsection{Generating the density maps}

A probability density is a function $f(s)$ defined over a region $A$
that allows the computation of the probability of locating an single
object in a subregion $B$ by integrating $f$ over the subregion $B$:
$\int_B f(s) ds = P({\rm object\ in\ B})$.  In particular, areas
where the density is high are areas where finding an object is
likely.  The minimal requirements of such a function are that $\int_R
f(s)ds = 1$, so that the probability of finding a specific object
(i.e. one specific fish) somewhere in the region is 1, and that $f(s)\geq 0$, to avoid negative probabilities.

Given a point process of object locations, the true density can be
estimated by means of a smoothing process, wherein areas in which
large numbers of objects are found in a restricted are given high
estimated densities.  A density estimator $\hat{f}$ should be a
continuous function over our region and should be a bona fide
density, that is its integral over the entire region should be one,
and the estimated density should be non-negative everywhere.

Our approach is to discretize $\hat{f}$ by choosing a set of $N$
nodes (locations) in the region and defining a probability at each
node such that the sum of the probabilities over all nodes is one.
The relationship between the density estimate $\hat{f}$, which is
defined everywhere over the region, and the probabilities at a node
$s_i$ is that the node probabilities approximate the integral of
$\hat{f}$ over a small region around $s_i$ with area $area(A)/N$. We
will compute the probabilities at $s_1,...,s_N$ from a random walk
on the nodes, then obtain the estimated density from $\hat{f}(s_i)$
equal to the (discretized) probability at $s_i$ divided by
$area(A)/N$. To find the discrete probability densities at
$s_1,...,s_N$ we need to define a lattice over these nodes.

A lattice consists of a set of $N$ nodes with NS and EW coordinates,
and a neighbor relationship between pairs of nodes (Cressie 1993,
pp. 383 ff).  The neighbor relationship may be defined in many ways.
Our implementation defines neighbors to be the closest nodes in the
N, S, E, W, NE, NW, SE and SW directions, although it is possible to
add links or remove them depending on the judgement of the
researcher, for a particular data set.  By definition the nodes are
also their own neighbors.

Our estimator is the probability density of the length-$k$ random
walk on the lattice.  Let $X_k$ denote the position of the random
walk at at step $k$.  When $k=0$ the estimated density is just the
original set of observations, so that $P(X_0 = s_i)= $ the
proportion of observed fish at location $s_i$. The random walk is
just a finite state Markov chain, familiar to anyone who has worked
with age-structured population models (Leslie 1945), with transition
probabilities $P(X_{k+1}=s_i | X_k = s_j)\not= 0$ only if $s_i$ is a
neighbor of $s_j$.  Following standard Markov chain notation, the
$N\time 1$ probability vector $p_k$ is

$$p_k = \lbrack P(X_k = s_1),\cdots P(X_k = s_N)\rbrack$$
which shows the probability distribution after $k$ steps.

Define $T$ to be a $N\times N$ transition matrix in which the entry
in the ith row and jth column is $P(X_{k+1} = s_j | X_k = s_i)$, the
probability that a random walk at location $s_i$ moves to
neighboring location $s_j$.  From basic probability theory, $p_{k+1}
= Tp_k$, so that multiplying the probability density at time $k$ by
$T$ produces the probability density after one step.  By repeating
this process, $p_k = T^kp_0$ gives the probability density of the
random process after $k$ steps.

\subsection{Properties of the transition matrix}

For purposes of density estimation, especially mimicking the
behavior of the usual kernel density estimator, it would be
desirable that as $k$ gets very large, the density converges to the
uniform (flat) density where $P(X_k = s_i)\approx 1/N$ at all
locations when these locations are connected, since the ultimate
smoothed estimator is constant everywhere in a connected region.
Also, when there is no boundary and $k$ is moderately large, the
density should be approximately the sum of normal kernels centered
at each observation, so that in the no-boundary case the
lattice-based estimator is comparable to the usual kernel density
estimator.

In the no-boundary case, $X_k$ is a symmetric random walk centered
on $X_0$, thus is the sum of $k$ independent and identically
distributed random variables with finite variances, so that the
central limit theorem makes the density approximate a bivariate
normal density for moderate to large $k$.

A sufficient condition for the random process to ultimately become
uniform is that $T$ be symmetric and all of the nodes connected in
the sense that we can get from one node to any other by moving from
node to neighboring node (Rosenblatt 1971). We add the further
assumption that all movement probabilities are the same, so that the
rate of movement of the random walk is the same everywhere. This is
guaranteed by choosing the transition probabilities as follows:
 First, define $q_i$ to be the number of neighbors (other than itself) of the location
 $s_i$ (usually the maximum value for $q_i$ is 8 when a location is
 not near a boundary).  Next, chose a parameter $M$ between zero and
 one.  This parameter governs how often the random walk remains in the same
 location after one step.  Then the transition probabilities are

\begin{equation}
P(X_{k+1} = s_i | X_k = s_i) = 1 - M*(q_i/\max(q_i))
\label{eq:transprob1}
\end{equation}

and

\begin{equation}
P(X_{k+1} = s_j | X_k = s_i) = M*(1/\max(q_i))\ \ {\rm for}\ \ i\not= j
\label{eq:transprob2}
\end{equation}

In this paper we use $M = 0.5$. Generally the higher $M$ is, the
more steps will be required to achieve the same degree of smoothing.
The process as described above is a valid transition matrix  and
symmetric.  To find the probabilities on $s_1,...,s_N$, it remains
only to choose the number of steps $k$. Then the estimated density
is

\begin{equation}
p_k = T^kp_0,\quad \hat{f}(s_i) = (N/area(A))p_{k,i}
\label{eq:estden}
\end{equation}

\subsection{Crossvalidation}

As the number of steps $k$ increases, the resulting density map
becomes smoother. Selecting the optimal value of $k$ is analogous to
selecting the bandwidth in kernel density estimation. We suggest a
crossvalidation approach. Here, for each observed fish or animal at
location $s_i$, we start with a probability density $p_{0, -i}$
giving weight $1/(n-1)$ to all other observations and removing the
ith observation.  At step $k$, define $p_{k,i,-i}$ as the ith
element in $T^kp_{0, -i}$.  What we are doing is removing the ith
observation, then determining the density at location $s_i$ after
$k$ steps. These are then combined into a measure of goodness-of-fit
at k steps, the Unbiased Crossvalidation criterion UCV (Sain,
Baggerly, Scott 1994):

$$UCV_k = \int_{R^2} \hat{f}^2(x) dx - {2\over n}\sum_{i=1}^n \hat{f}_{-i}(x_i)
$$
which we approximate by
\begin{equation}
UCV_k = {N\over area(A)}\sum_{j=1}^N p_j^2 - {N\over area(A)}{2\over
n}\sum_{i=1}^n p_{k,i,-i} \label{eq:ucv}
\end{equation}
where $N$ is the number of nodes, $n$ is the number of observation
and $area(A)$ is the area of the region.  The optimal number of
steps is that which minimizes $UCV_k$.

\subsection{Simple example}

\begin{figure}
 \centerline{\includegraphics[width=3in]{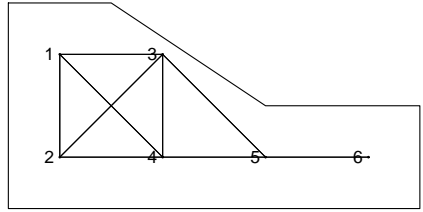}}
 \renewcommand{\baselinestretch}{1} \caption{Example lattice consisting of six nodes and nine bidirectional links inside a polygonal region.}
\label{fig:fig1}
\end{figure}

A very simple example should illustrate the mathematics of this
technique.  Figure (\ref{fig:fig1}) shows a polygon with a lattice
consisting of six nodes and nine bidirectional links. Recall that
$q_i$ is the number of neighbors (not counting itself) of location
$s_i$.  Here $q_1 = 3, q_2 = 3, q_3 = 3, q_4 = 4, q_5 = 3, q_6 = 1$.
With $M = 0.5$, application of equation (\ref{eq:transprob1}) yields
the transition matrix T:

$$
T =\left(
  \begin{array}{cccccc}
0.625 &0.125 &0.125 &0.125 &0.000 &0.000\\

0.125 &0.625 &0.125 &0.125 &0.000 &0.000\\

0.125 &0.125 &0.500 &0.125 &0.125 &0.000\\

0.125 &0.125 &0.125 &0.500 &0.125 &0.000\\

0.000 &0.000 &0.125 &0.125 &0.625 &0.125\\

0.000 &0.000 &0.000 &0.000 &0.125 &0.875\\
  \end{array}
\right)
$$

For instance this tells us that there is a 0.625 chance that a
random walk at location 1 remains at location 1 after a single step,
that there is a 0.125 chance that it moves from location 1 to
location 2, and that there is no chance that it moves from location
1 to location 5 in a single step.

\smallskip

Note that the probability of movement from one node to a neighboring
node is always $0.125$ in this example and movement directly to a
non-neighbor is impossible.  Figure \ref{fig:fig1} shows the lattice with nodes
labeled.

Suppose that one observation is recorded at location 1 and two
observations are recorded at location 3. Then the initial
probability density is $p_0 = \lbrack 0.3333, 0, 0.6667, 0, 0,
0\rbrack$.  After a single step the density is $p_1 = Tp_0 =$
$\lbrack 0.2916, 0.1250, 0.3750, 0.1250, 0.0833, 0.0000\rbrack$. One
step isn't sufficient, of course, to get non-zero probability at
node 6, but the density is no longer concentrated only on nodes 1
and 3.

After two steps the density is more dispersed, with probability $
p_2 =$ $ T^2p_0 =$ $\lbrack 0.2604, 0.1770, 0.2656,$ $ 0.1718,
0.1145, 0.0104\rbrack$.  After thirty steps the probability density
has become close to uniform, with probabilities $p_{30} = T^{30}p_0
=$ $\lbrack 0.1703, 0.1703, 0.1689, 0.1689, 0.1643, 0.1570\rbrack$.  This diffusion process is illustrated in Figure \ref{fig:fig2}.

\begin{figure}
 \centerline{\includegraphics[width=3in]{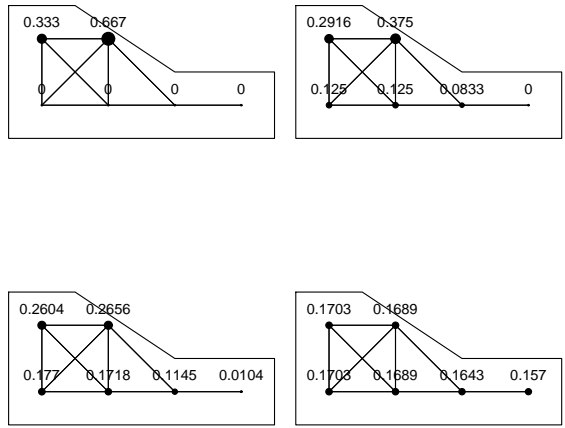}}
 \renewcommand{\baselinestretch}{1} \caption{Probability density of the random walk at zero, one, two and thirty steps in the simple example.}
\label{fig:fig2}
\end{figure}

\subsection{Another example}

The lattice-based density estimator performs well when there are
convoluted boundaries.  Consider the simulated point process displayed in
Figure 3.

\begin{figure}
 \centerline{\includegraphics[width=3in]{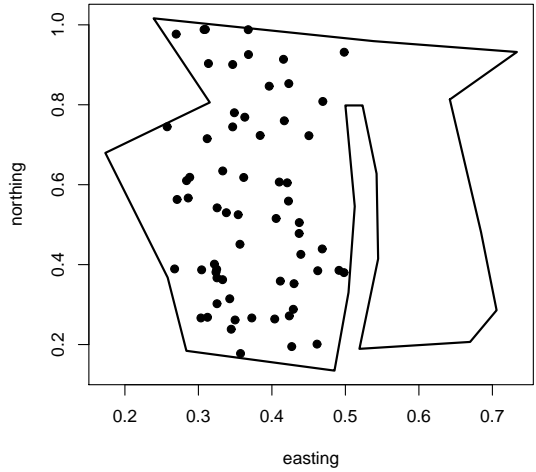}}
 \renewcommand{\baselinestretch}{1} \caption{A complicated polygon, with a point process restricted west of a causeway.}
\label{fig:fig2}
\end{figure}

The point process was generated  assuming a constant density for
easting values less than 0.5 (see Figure 3). Clearly estimating
density in this example with a standard approach that ignores the
boundary information will not be able to prevent density from
crossing the causeway and giving improper higher densities in areas
just to the east of the causeway. We estimated the density with the
lattice-based kernel estimator.  A set of nodes spaced 0.01 units
apart was juxtaposed over the polygon.  Crossvalidation yielded a
minimum UCV at k = 176 steps. The large number of steps is
reasonable, since we would expect the map to be quite smoothed since
the true density is uniform over much of the polygon.  The resulting
density map is shown in Figure 4.

\begin{figure}
 \centerline{\includegraphics[width=3in]{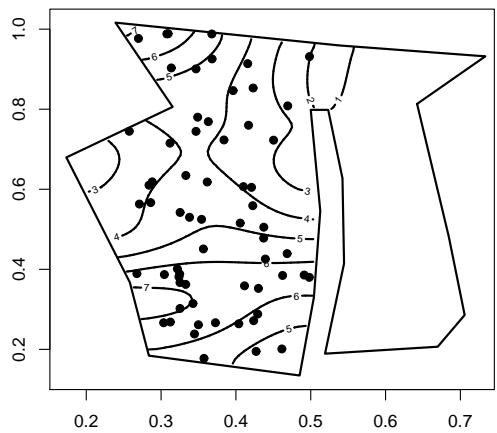}}
 \renewcommand{\baselinestretch}{1} \caption{Lattice-based density estimate based on k = 176 steps.}
\label{fig:fig2}
\end{figure}

Note that the density estimate is positive in the eastern side of
the polygon where there is an opening in the causeway, but not in
other areas east of the causeway.

\section{Simulation}

We performed a small simulation study to compare the performance of
our estimator to that of the usual bivariate kernel density
estimator.  Since the typical kernel estimator does not account for
irregular boundaries or holes, we computed both estimators over a
square region.  Thus this simulation only considers the situation where
the boundary does not matter.  However the results can be used to
infer the performance of our estimator over a more complex region.

In each simulation we generated $n=100$ observations, $(X_j, Y_j)$
for $j=1, \ldots, 100$, from the multivariate normal distribution
with mean and covariance
$$
\bmu = \left[\begin{array}{c} 5 \\ 5 \end{array}\right] \qquad\hbox{and}\qquad \bSigma =
 \left[ \begin{array}{cc} 1.5 & 0.8 \\0.8 & 1.5 \end{array}\right].
$$
Data were generated using the R package {\tt mnormt} (Genz and Azzalini 2009).

For each simulated data set we computed the lattice-based density
estimator in Equation (\ref{eq:estden})
with $k$ chosen by minimizing the UCV criterion in (\ref{eq:ucv}),
and the bivariate kernel density estimator (Venables and Ripley
2002),
$$
\wh f(x, y) = \frac{1}{nh_1h_2}\sum_{j=1}^n \phi\left(\frac{x-X_j}{h_1}\right)\phi\left(\frac{y-Y_j}{h_2}\right),
$$
where $\phi$ is the Gaussian kernel and $h_1$ and $h_2$ are
bandwidth parameters.  Bandwidths for the kernel estimator were also
chosen by Unbiased Crossvalidation (Venables and Ripley 2002).  This
estimator and its bandwidth were computed with the R functions {\tt
kde2d} and {\tt ucv} in the package {\tt MASS} (Venables and Ripley
2002).  Both estimators were computed on a $16\times 16$ grid from
-10 to 10 in both $x$ and $y$ directions.

Estimators were compared based on their average Integrated Squared
Error (ISE), defined for an estimator $\wh f(x,y)$ as
$$
ISE\{\wh f(x, y)\} = \int\int \left\{f(x, y) - \wh f(x, y)\right\}^2dxdy.
$$
Results are based on 100 simulated data sets.

\begin{figure}
 \centerline{\includegraphics[width=6in,height=2in]{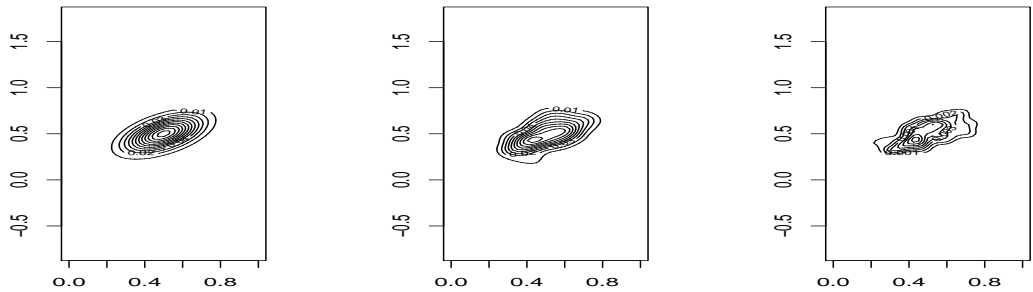}}
 \renewcommand{\baselinestretch}{1} \caption{The true density, a kernel density estimate and a lattice-based density estimate}
\label{fig:simfig}
\end{figure}

\begin{figure}
 \centerline{\includegraphics[width=3in]{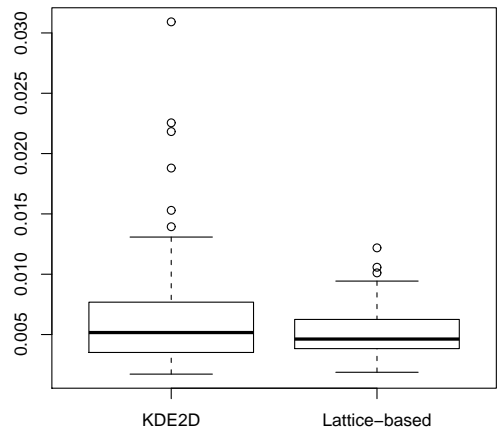}}
 \renewcommand{\baselinestretch}{1} \caption{Boxplots of average ISE for the lattice-based and kernel estimators.}
\label{fig:ise}
\end{figure}

The normal target density, along with one realization of the
bivariate kernel and lattice-based density estimators, are displayed
in Figure \ref{fig:simfig}.  In terms of average ISE, the
lattice-based density estimator actually performed significantly
better then the two-dimensional kernel density estimator when
compared using a paired t-test (p-value = 0.00169). There was also
less variation in ISE for the lattice-based density estimator than
the kernel estimator.  Boxplots of the ISEs are displayed in Figure
\ref{fig:ise}.

\section{Application}

\begin{figure}
 \centerline{\includegraphics[width=5in]{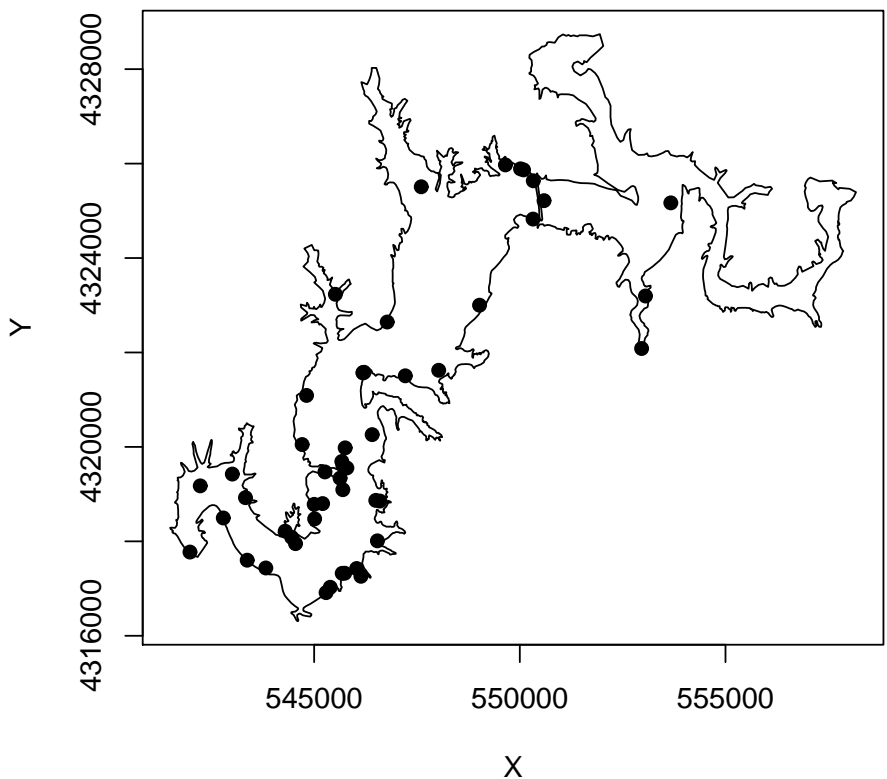}}
 \renewcommand{\baselinestretch}{1} \caption{Relocations of walleye in Lake Monroe.}
\label{fig:lake}
\end{figure}

We illustrate our method using data supplied by Sandra Clark-Kolaks
based on radiotelemetry relocations of walleye ({\it Sander
vitreus}) in Lake Monroe, Indiana (Clark-Kolaks 2009). Data from
three widely-separated days, 16 October 2008 (15 relocations), 17
March 2009 (19 relocations) and 9 April 2009 (16 relocations), were
pooled for at total of 50 relocations. These relocations are plotted
within the polygonal representation of Lake Monroe in Figure
\ref{fig:lake}.  It is clear that the standard kernel approaches
will be problematic in estimating walleye home range in the lake.
Many of the relocations are along the boundary of the lake, there is
a narrow causeway almost bisecting the lake and some of the fish
have been located in narrow bays.

We used the lattice-based density estimator to estimate walleye home
range. Nodes were spaced 200 m apart within the polygon, to cover
the entire lake.  The lattice was constructed in two steps.  First,
all pairs of nodes between 100 m and 300 m were declared neighbors.
Often this is sufficient, but in this case the convoluted nature of
the lake shore required some editing of the neighbor structure.
Three nodes had no neighbors at all and some pairs of nodes that
were separated by land were neighbors.  An editing function in R,
{\tt edit.lattice}, was used to hand-edit the neighbor structure.

\begin{figure}
 \centerline{\includegraphics[width=5in]{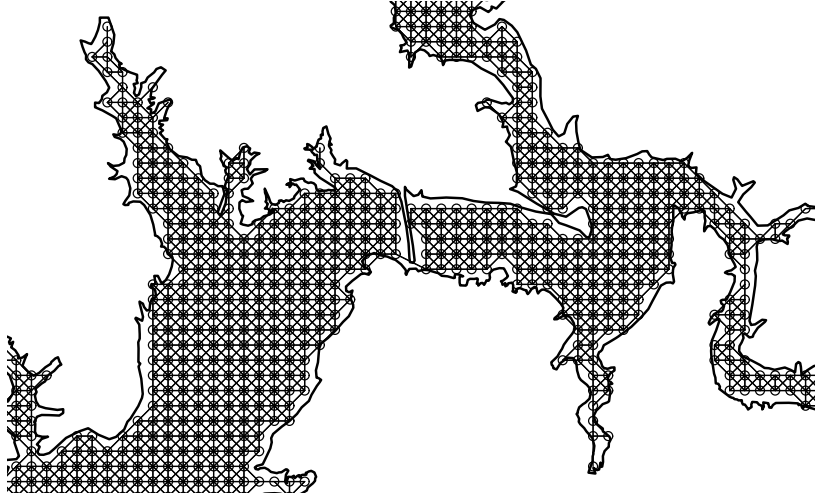}}
 \renewcommand{\baselinestretch}{1} \caption{View of the lattice and neighbor relationship around the causeway.}
\label{fig:lattice2}
\end{figure}

Figure \ref{fig:lattice2} shows a close look at the causeway and illustrates the
neighbor relationship, where neighbors are connected by line
segments.

\begin{figure}
 \centerline{\includegraphics[width=5in]{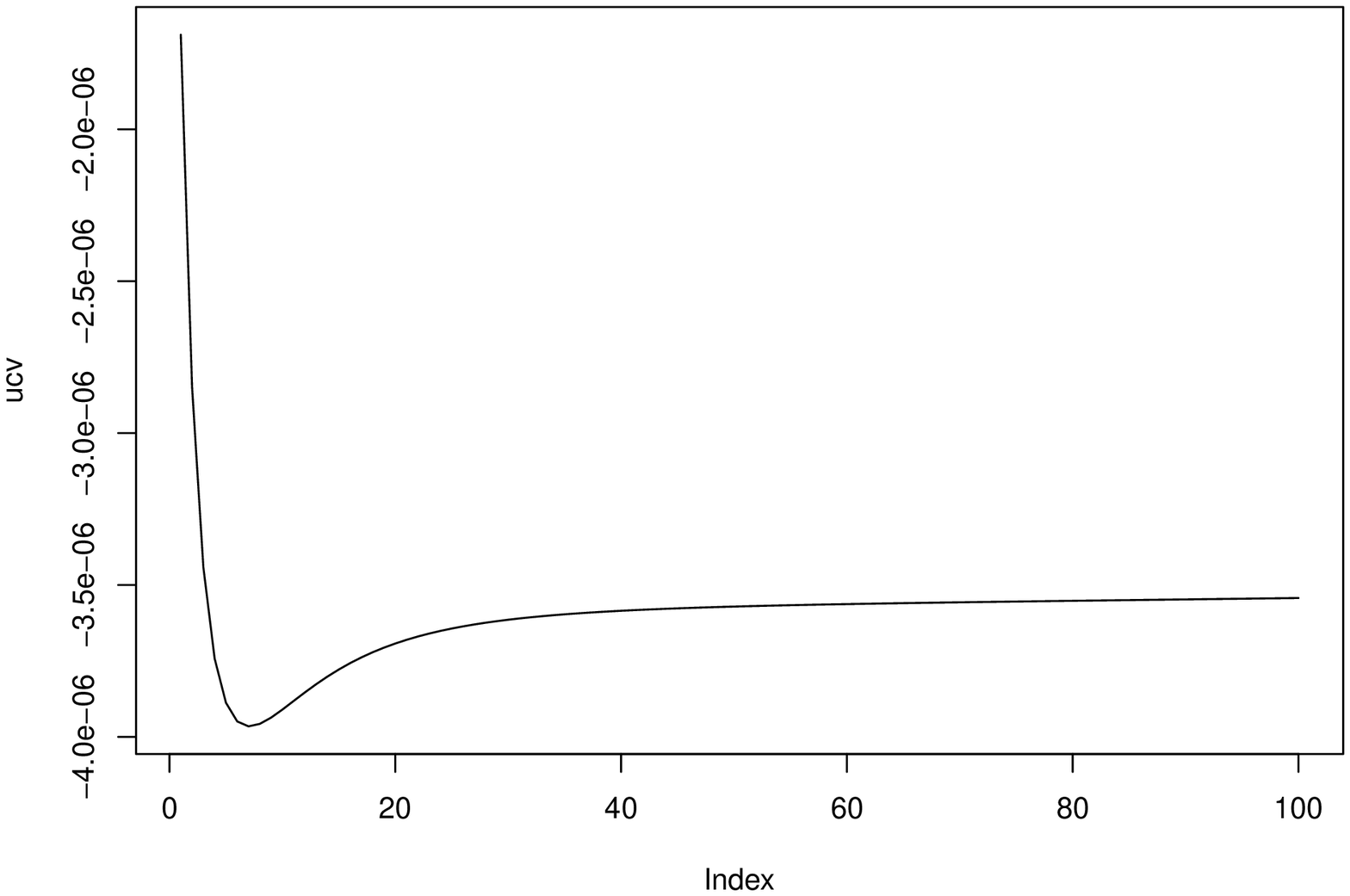}}
 \renewcommand{\baselinestretch}{1} \caption{Crossvalidation criterion for different numbers of steps $k$.}
\label{fig:ucvplot}
\end{figure}

\begin{figure}
 \centerline{\includegraphics[width=5in]{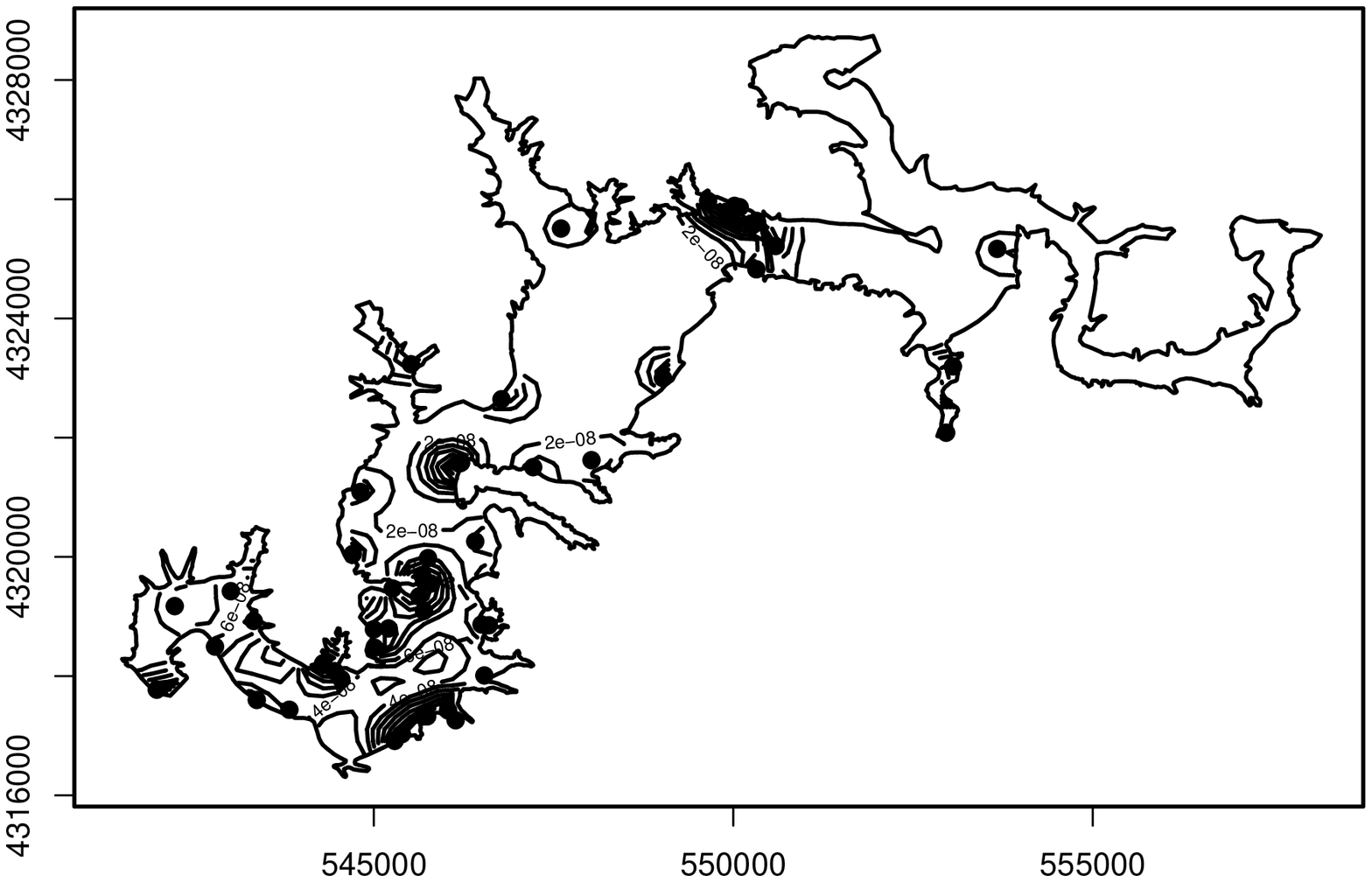}}
 \renewcommand{\baselinestretch}{1} \caption{Contour map of the lattice-based estimator of walleye home range in Lake Monroe.}
\label{fig:estimate1}
\end{figure}

Crossvalidation  on the data resulted in  a minimum UCV at 7 steps.
The approximate UCV at different values of $k$ is displayed in
Figure \ref{fig:ucvplot}.  The lattice-based density estimator was
computed with this number of steps, and a contour map of the
resulting estimate is shown in Figure \ref{fig:estimate1}.  Note
that the lattice-based density estimator shows higher densities in
narrow bays in which fish were observed.

\begin{figure}
 \centerline{\includegraphics[width=5in]{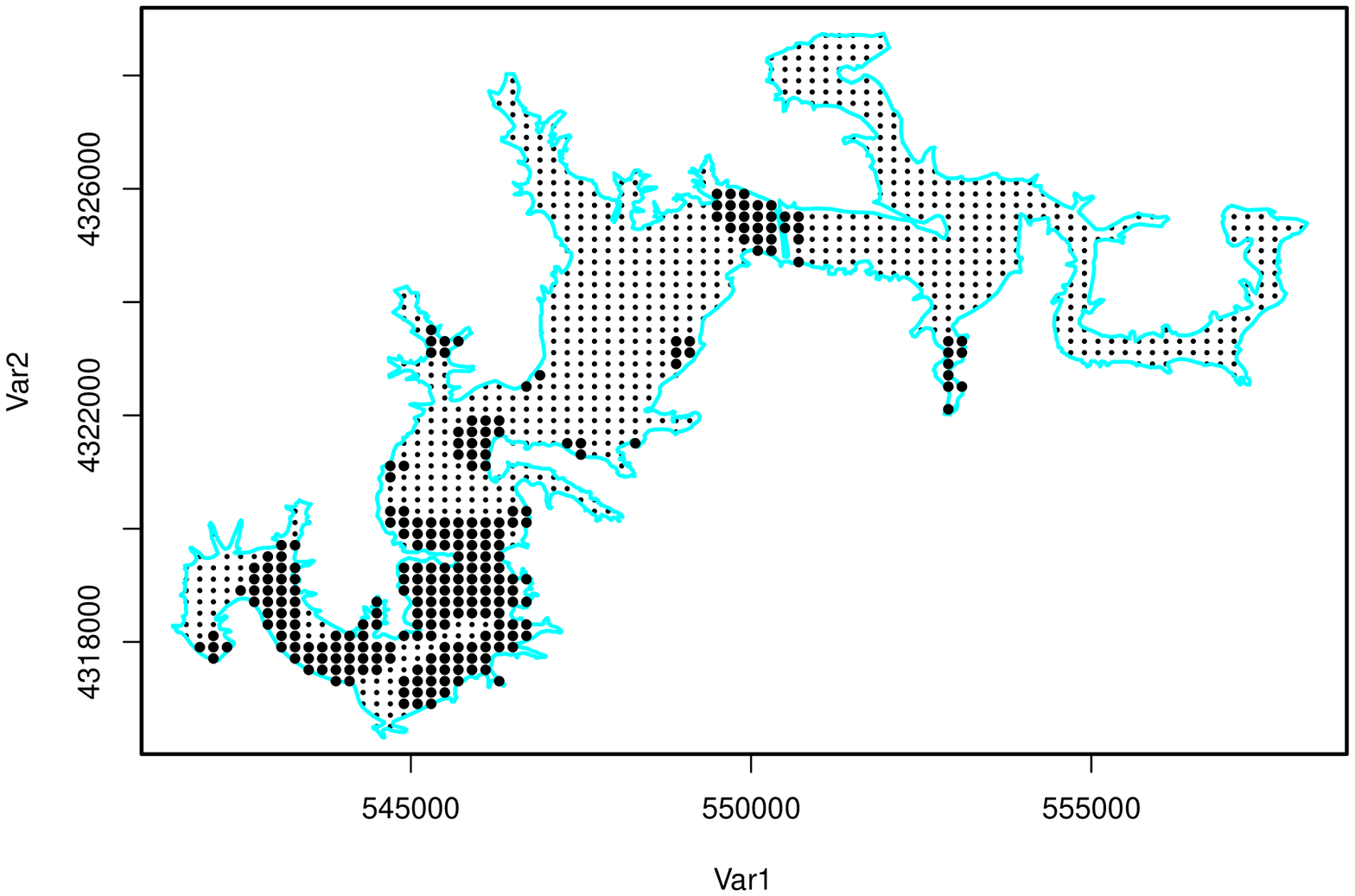}}
 \renewcommand{\baselinestretch}{1} \caption{Minimal area subregion accounting for at least probability 0.75}
\label{fig:estimate2}
\end{figure}

A common approach to defining a home range is to find the smallest
area that contains a given proportion P of the density, i.e. a
subregion B of minimal area such that $\int_B \hat{f}(s)ds = P$.
 We find such a region simply by finding the smallest number of nodes that
 have a probability totalling at least $P$.  Figure \ref{fig:estimate2} shows the
 minimal area where the nodes
account for at least $P = 0.75$.

\section{Discussion}

The presence of boundaries and holes in regions has been problem for
the typical implementation of kernel density estimators for home
range or utilization distribution studies.  The lattice-based
density estimator adjusts densities for boundaries and holes, and is
as implemented in R fairly straightforward to use.  In the absence
of boundaries and with elliptical true densities the lattice-based
estimator works as well as the typical kernel density estimator so
there is no reason not to use the lattice-based approach. Because
bandwidth can be obtained through crossvalidation, the only decision
required of the researcher is deciding on the spacing of the nodes,
which realistically should be as small as possible, limited by the
computation time and memory required.  For the Lake Monroe data, on
a fairly antiquated laptop computer (Dell Inspiron 9300, running
Windows XP at 1.6 GHz with 2 GB of RAM) the generation of the nodes
and lattice required 4 seconds while the crossvalidation required
only 8 additional seconds.  However, reducing the node spacing down
from 200 meters to 50 meters made the functions run very slowly as
too much memory was required.  All of the R functions are available
as an R package from the authors.

\section{Acknowledgements}

We thank Sandra Clark-Kolaks for the use of the Lake Monroe fish
relocations and shapefiles, along with the Indiana Department of
Natural Resources which supported the Lake Monroe data collection.
Our analysis of this data is for illustrative purposes only and has
not been vetted by the Indiana Dept. of Natural Resources.

\bigskip

{\baselineskip = 16pt \parindent=0pt

REFERENCES:

\begin{list}{}{}

\item Clark-Kolaks S.  Distribution and movement of walleye (Sander
vitreus) in Monroe Reservoir, Indiana 2008 and 2009,  Fish Research
Final Report.  [internet]  Indiana:  Fisheries Section, Indiana
Department of Natural Resources, Division of Fish and Wildlife; 2009
[cited 30 Sept 2010] Available from: http://www.in.gov/dnr/
fishwild/files/
fw-Monroe\_Reservoir\_Walleye\_Movement\_2008\_and\_2009.pdf
\medskip

\item Cressie NA.  Statistics for Spatial Data:  Revised Edition. New
York: John Wiley and Sons, Inc.; 1993
\medskip

\item Downs JA and Horner MW.  Network-Based Kernel Density Estimation
for Home Range Analysis. In: Proceedings of the 9th International
Conference on Geocomputation 2007 (Maynooth, Ireland), National
University of Ireland Maynooth: Maynooth, Ireland; 2007
\medskip

\item  Genz A and Azzalini A.
  mnormt: The multivariate normal and t distributions. R package
  version 1.3-3. http://CRAN.R-project.org/package=mnormt; 2009.
\medskip

\item  Getz WM,  Fortmann-Roe S, Cross PC, Lyons AJ, Ryan
  SJ, Wilmers CC.  LoCoH: Nonparametric Kernel Methods for
  Constructing Home Ranges and Utilization Distributions.  PLos ONE
  2(2):  e207.  doi:10.1371/journal.pone.0000207; 2009.
\medskip

\item  Getz WM, Wilmer CC. A local nearest-neighbor convex-hull
  construction of home ranges and utilization distributions.
  Ecography 27: pp. 489-505; 2004.
\medskip

\item  Karunamuni RJ and Alberts T On boundary correction in kernel
density
  estimations.  Statistical Methodology
2 (3) pp. 191-212; 2005.
\medskip

\item R Development Core Team. R: A language and environment for
  statistical computing. R Foundation for Statistical Computing [online],
 Vienna, Austria. ISBN 3-900051-07-0, Available from
http://www.R-project.org.; 2009.
\medskip

\item  Ryan SJ, Knechtel CU and Getz WM.  Range and habitat
  selection of African buffalo in South Africa.  J Wildlife
  Management 70:  pp. 764-776; 2006.
\medskip

\item Sain SR, Baggerly KA, Scott DW.  Cross-validation of
multivariate densities.  JASA.  89 (427) pp. 807 - 817; 1994.
\medskip

\item  Venables WN and Ripley BD. Modern Applied Statistics with
  S. Fourth Edition. Springer, New York. ISBN 0-387-95457-0; 2002.
\medskip

\item Worton BJ.  Kernel methods for estimating the utilization
distribution in home-range studies.  Ecology.  70 (1)  pp. 164 -
168; 1989.

\end{list}

}

\end{document}